\documentclass[aps,prl,twocolumn,groupedaddress,showpacs,showkeys,amsmath,amssymb]{revtex4}
\usepackage{graphicx}
\begin{document}
\title{High-field vortices in Josephson junctions with alternating
critical current density}
\author{M. Moshe}
\affiliation{School of Physics and Astronomy, Raymond and Beverly Sackler
Faculty of Exact Sciences, Tel Aviv University, Tel Aviv 69978, Israel}
\author{R. G. Mints}
\email[]{mints@post.tau.ac.il}
\affiliation{School of Physics and Astronomy, Raymond and Beverly Sackler
Faculty of Exact Sciences, Tel Aviv University, Tel Aviv 69978, Israel}
\par
\date{\today}
\begin{abstract}
We study long Josephson junctions with the critical current density
alternating along the junction. New equilibrium states, which we call
the field synchronized or FS states, are shown to exist if the applied
field is from narrow intervals centered around equidistant series of
resonant fields, $H_m$. The values of $H_m$ are much higher than the
flux penetration field, $H_s$. The flux per period of the alternating
critical current density, $\phi_i$, is fixed for each of the FS states.
In the $m$-th FS state the value of $\phi_i$ is equal to an integer
amount of flux quanta, $\phi_i =m\phi_0$. Two types of single Josephson
vortices carrying fluxes $\phi_0$ or/and $\phi_0/2$ can exist in the FS
states. Specific stepwise resonances in the current-voltage
characteristics are caused by periodic motion of these vortices between
the edges of the junction.
\end{abstract}
\maketitle
\begin{figure}
\includegraphics[width=0.95\columnwidth]{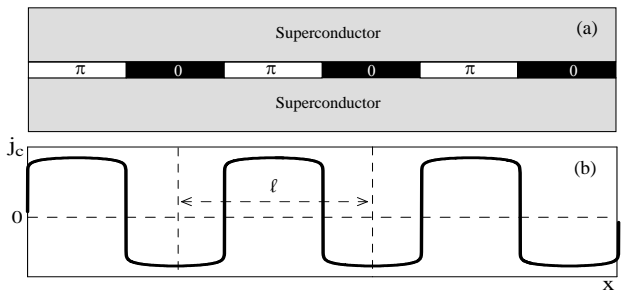}
\caption{Schematic diagrams of: (a) tunnel junction arranged in
series of interchanging $0\,$- and $\pi\,$-shifted fragments; (b)
spatial distribution of alternating critical current density.
\label{fig_1}}
\end{figure}

Josephson tunnel structures arranged in a sequence of interchanging
$0\,$- and $\pi\,$-$\,$shifted fragments (see Fig.~\ref{fig_1}) are a
subject of growing interest \cite{Bulaevskii_1, Buzdin_1, Ryazanov_1,
Kontos_1, Blum_1, Copetti_1, Hilgenkamp_1, Harlingen_1, Tsuei_1,
Hilgenkamp_2, Mannhart_1, Mints_1, Mints_2, Mints_3, Buzdin_2}.
Currently, the properties of these complex systems are thoroughly
treated for superconductor-ferromagnet-superconductor (SFS)
heterostructures \cite{Bulaevskii_1, Buzdin_1, Ryazanov_1, Kontos_1,
Blum_1} and asymmetric grain boundaries in thin films of high-T$_c$
superconductor YBCO \cite{Copetti_1, Hilgenkamp_1, Harlingen_1,
Tsuei_1, Hilgenkamp_2, Mannhart_1, Mints_1, Mints_2, Mints_3,
Buzdin_2}.
\par
It has been predicted that in the SFS Josephson junctions the
$\pi$-shift in the phase difference $\varphi$ between the
superconducting banks is caused by the ferromagnet interlayer
\cite{Bulaevskii_1, Buzdin_1}. This prediction was confirmed in recent
experiments with SFS $\pi$-$\,$shifted junctions and SFS
heterostructures of interchanging $0\,$- and $\pi\,$-$\,$shifted
fragments \cite{Ryazanov_1, Kontos_1, Blum_1}. The asymmetric grain
boundaries in YBCO thin films are arranged in series of facets with
variety of orientations and lengths $l\sim 10 - 100\,$nm
\cite{Hilgenkamp_2}. This spatial structure in conjunction with the
$d$-wave symmetry of the order parameter results in grain boundary
junctions with interchanging $0\,$- and $\pi\,$-$\,$shifted fragments
\cite{Harlingen_1, Tsuei_1}.
\par
% fig. 1
\par
The critical current density, $j_c(x)$, changes sign at each contact
between $0\,$- and $\pi\,$-$\,$shifted fragments ($x$ axis is along the
junction). This results in a variety of dramatic changes in Josephson
properties. in particular, the field dependence of the maximum
supercurrent across the junction, $I_{\rm m}$, is strongly affected by
the alternating $j_c(x)$ \cite{Copetti_1, Hilgenkamp_1, Mints_1,
Buzdin_2}. First, the value $I_m(H_a)$ is significantly suppressed at
low fields, $|H_a|\ll H_1=\phi_0/ 2\lambda l$, where $\lambda$ is the
London penetration depth and $l$ is the period of $j_c(x)$. Second,
unlike the standard Fraunhofer pattern with a major peak at $H_a=0$,
two major side-peaks are observed at high fields, $H_a=\pm H_1$, where
$H_1\gg H_s$ and $H_s$ is the field of first flux penetration
\cite{Hilgenkamp_1}.
\par
In this letter we find a series of new equilibrium {\it field
synchronized} (FS) states existing if the applied field is from narrow
intervals $\Delta H_a\sim H_s$ centered at the resonant fields $H_m=\pm
mH_1$, where $m\ne 0$ is an integer. It is shown that in the $m$-th FS
state the flux per period of the alternating critical current density,
$\phi_i$, is fixed and is equal to an integer number of flux quanta,
$\phi_i=m\phi_0$. We demonstrated that two types of high-field ($H_a\gg
H_s$) Josephson vortices with fluxes $\phi_0$ or/and $\phi_0/2$ can
exist in the FS states. We find that periodic motion of these single
high-field vortices cause stepwise resonances in the IV curves similar
to the zero-field resonances in Josephson junctions of conventional
superconductors \cite{Tinkham_1}.
\par
We begin with a qualitative treatment of the FS states using one
harmonics model for the tunneling current density
$j=j_c(x)\sin\varphi$, where $j_c(x)=j_1\sin(2\pi x/l)$, $L=Nl$, $L$ is
the length of the junction, and $N\gg 1$ is an integer. Assume that the
junction is in one of the FS states and flux $\phi_i$ is fixed. Since
we have many vortices in the junction ($N\cdot m\gg 1$), the field is
almost uniform and the phase $\varphi (x)$ takes the form
% 01
\begin{equation}\label{eq_001}
\varphi (x)=2\pi{\phi_i\over\phi_0}{x\over l} +\psi (x)\,,
\end{equation}
where the phase $\psi(x)$ is a smooth function with the typical length
scale $\gg l$ and $|\psi (x)|\sim 1$. Then we have:
% 02
\begin{equation}\label{eq_002}
j(x)=j_1\sin\left(2\pi{x\over
l}\right)\sin\left(2\pi{\phi_i\over\phi_0}{x\over l} +\psi\right)\,.
\end{equation}
In general, this current density alternates rapidly with a typical
length scale $\le l$. In this case the coarse-grained approach is the
right tool to describe the smooth phase $\psi (x)$ \cite{Mints_2}. If
we average $j(x)$ over a distance ${\cal L}\gg l$, then the
coarse-grained tunneling current density, $j_\psi$, is zero. This is
indeed true in all cases but one. If the junction is in the FS state
with $\phi_i=\pm\phi_0$, then coarse-graining of Eq. (\ref{eq_002})
leads to a nonzero result
% 03
\begin{equation}\label{eq_003}
j_\psi = 0.5\,j_1\sin (\psi\pm\pi/2).
\end{equation}
\par
It is worth mentioning that $j_\psi\sim j_1$ and the final form of the
dependence of $j_\psi$ on the smooth phase $\psi$ coincides with a
$\pi/2\,$-shifted current-phase relation of a Josephson junction of
conventional superconductors.
\par
Similar calculation of the coarse-grained current density $j_\psi$ in
the low-field region ($H_a\ll H_s$) leads to \cite{Mints_2}
% 04
\begin{equation}\label{eq_004}
j_\psi =-j_1\left({l\over 4\pi\Lambda_1}\right)^2\sin 2\psi\ll j_1,
\end{equation}
where the Josephson length is given by
% 05
\begin{equation}\label{eq_005}
\Lambda_1=\sqrt{{c\phi_0/16\pi^2\lambda j_1}}\gg l\,.
\end{equation}
\par
\begin{figure}
\includegraphics[width=0.95\columnwidth]{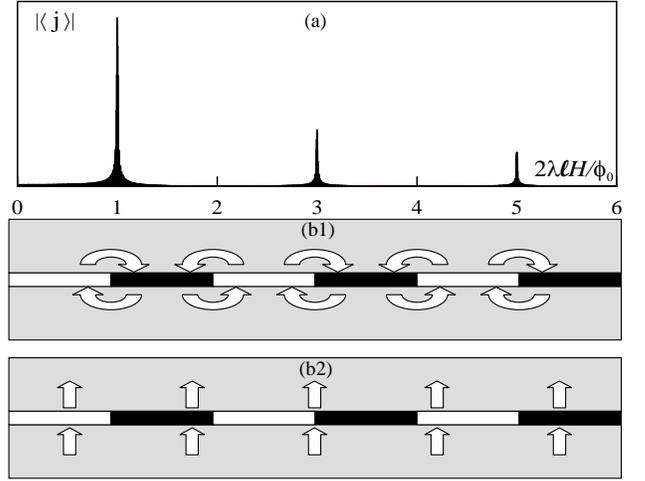}
\caption{(a) Coarse-grained tunneling current density dependence on
the applied field for the case of a stepwise critical current density,
$N=40$, and $\varphi\propto x$. Schematic diagrams of spatial
distributions of tunneling current density for junctions that are: (b1)
not in the FS states (small current loops tend to cancel each other);
(b2) in the FS states (field synchronized local currents flow in the
same direction).
\label{fig_2}}
\end{figure}

Comparing Eqs. (\ref{eq_002}), (\ref{eq_003}), and (\ref{eq_004}) we
find that if the phase factor, $\sin\varphi$, modulations caused by the
field and the critical current density modulations caused by the
structure of the junction are {\it synchronized} ($\phi_i=m\phi_0$),
then the critical current density $j_c(x)$ is significantly enhanced
(see Fig.~\ref{fig_2}). The widths of the intervals $\Delta H_a\sim
H_s\ll H_1$ of existence of the FS states are defined by minimization
of the free energy of the junction.
\par
Consider now a Josephson junction with $\lambda\ll l\ll\lambda_J$,
where $\lambda_J$ is the {\it local} Josephson penetration depth
% 06
\begin{equation}\label{eq_006}
\lambda_J=\sqrt{c\phi_0/16\pi^2\lambda \langle|j_c|\rangle},
\end{equation}
and the averaging over the junction length is defined as
% 07
\begin{equation}\label{eq_007}
\langle f\rangle =\int_0^L\!\!\! dx\, f(x)/L\,.
\end{equation}
\par
% fig. 2
\par
In this case the static spatial distribution of the phase $\varphi (x)$
is given by
% 08
\begin{equation}\label{eq_008}
\lambda_J^2\varphi''-i_c(x)\sin\varphi =0\,,
\end{equation}
where $i_c(x)=j_c(x)/\langle|j_c|\rangle$ is the dimensionless critical
current density. Next, we expend $i_c(x)$ in Fourier series and obtain
instead of Eq. (\ref{eq_008}):
% 09
\begin{equation}\label{eq_009}
\lambda_J^2\varphi''
-\sum_{n=-\infty}^{\infty}i_n e^{i2\pi nx/l}\sin\varphi =0\,,
\end{equation}
where $i_n$ are the Fourier transforms of the function $i_c(x)$.
\par
Since $l\ll\lambda_J$, the phase $\varphi$ can be written as
\cite{Arnold_1,Mints_2}
% 10
\begin{equation}\label{eq_010}
\varphi = 2\pi\phi_i x/\phi_0 l +\psi(x) + \xi(x)\,,
\end{equation}
where $\psi(x)$ is a smooth function with the typical length scale
$\sim\lambda_J\gg l$ and $|\psi (x)|\sim 1$, $\xi (x)$ is a rapidly
alternating function with the length scale $\sim l$. In addition, we
assume that $\langle\xi(x)\rangle =0$, and $\langle|\xi(x)|\rangle\ll
1$. Following \cite{Mints_2,Arnold_1} we average Eq. (\ref{eq_009})
over the junction length and obtain the equation describing the smooth
(coarse-grained) phase $\psi(x)$ in the $m$-th FS state
% 11
\begin{equation}\label{eq_011}
\lambda_J^2\psi'' =|i_m|\sin(\psi-\theta_m) -
\gamma_m\sin 2(\psi-\alpha_m),
\end{equation}
where $\theta_m=\arg(i_m)$, $\alpha_m$ and $\gamma_m$ are defined by
% 12
\begin{equation}\label{eq_012}
\gamma_m\,e^{-i2\alpha_m} =\left({l\over 2\pi\lambda_J}\right)^2
\,\sum_{n=1}^\infty {i_{m+n}i_{m-n}\over n^2}.
\end{equation}
\par
In general, complexity of Eq. (\ref{eq_011}) can be significantly
reduced. Indeed, one can estimate $\gamma_m\sim (l/2\pi\lambda_J)^2\ll
1$, which means that $\gamma_m\ll i_n$. As a result the second term in
the RHS of Eq. (\ref{eq_011}) can be neglected and Eq. (\ref{eq_011})
yields the sine-Gordon equation
% 13
\begin{equation}\label{eq_013}
\Lambda_m^2\psi'' =\sin(\psi-\theta_m)\,,
\end{equation}
where $\Lambda_m =\lambda_J/\sqrt{|i_m|}$. The boundary conditions to
Eq. (\ref{eq_013}) are given by the field at $x=0$ and $x=L$:
% 14
\begin{equation}\label{eq_014}
\psi'\big|_{\rm 0,L}={4\pi\lambda\over\phi_0}
\left(H\big|_{\rm 0,L}-mH_1\right).
\end{equation}
It is worth noting that, in particular, Eq. (\ref{eq_013}) describes
Josephson-type vortices with size $\sim\Lambda_m\gg l$ and flux
$\phi_0$.
\par
It follows from Eqs. (\ref{eq_011}) and (\ref{eq_013}) that equations
describing the coarse-grained phase $\psi(x)$ are the same as for
Josephson junctions of conventional superconductors in the Meissner
state.
\par
Specific symmetry of $j_c(x)$ might lead to a Fourier series with some
of the transforms being zero. In this case only the second term at the
RHS of Eq. (\ref{eq_011}) is non zero. In particular, if $j_c(x)$ is a
stepwise function (see Fig. \ref{fig_1}), then we have $i_{\rm 2k}=0$
and $i_{\rm 2k+1}=-2i/\pi\,(2k+1)$, where $k$ is an integer. In the FS
states with the field $H_a$ located in the intervals $\Delta H_a$
centered at ``odd'' resonant fields $H_{\rm ok} = (2k+1)H_1$, Eq.
(\ref{eq_011}) takes the form
% 15
\begin{equation}\label{eq_015}
\Lambda_{\rm ok}^2\psi''=\sin\psi, \quad \Lambda_{\rm ok}=
\sqrt{\pi (k+1/2)}\,\lambda_J\,.
\end{equation}
Similarly, for ``even'' resonant fields $H_{\rm ek} = 2kH_1$,
 Eq. (\ref{eq_011}) yields
% 16
\begin{equation}\label{eq_016}
\Lambda_{\rm ek}^2\psi''=\sin 2\psi, \quad \Lambda_{\rm ek} =
2\pi k\beta_k\,{\lambda_J^2\over l}\,,
\end{equation}
where $\beta_k$ is a constant ($\beta_1 =0.9$, $\beta_2 =1.1$, $\beta_3
=0.9$, and $\beta_k\approx 1$ for $k>3$). It is seen from Eqs.
(\ref{eq_015}) and (\ref{eq_016}) that the ``even''-field vortices
carrying flux $\phi_0/2$ are by the factor $\lambda_J/ l\gg 1$ wider
than the ``odd''-field vortices carrying flux $\phi_0$.
\par
% fig. 3
\begin{figure}
\includegraphics[width=0.95\columnwidth]{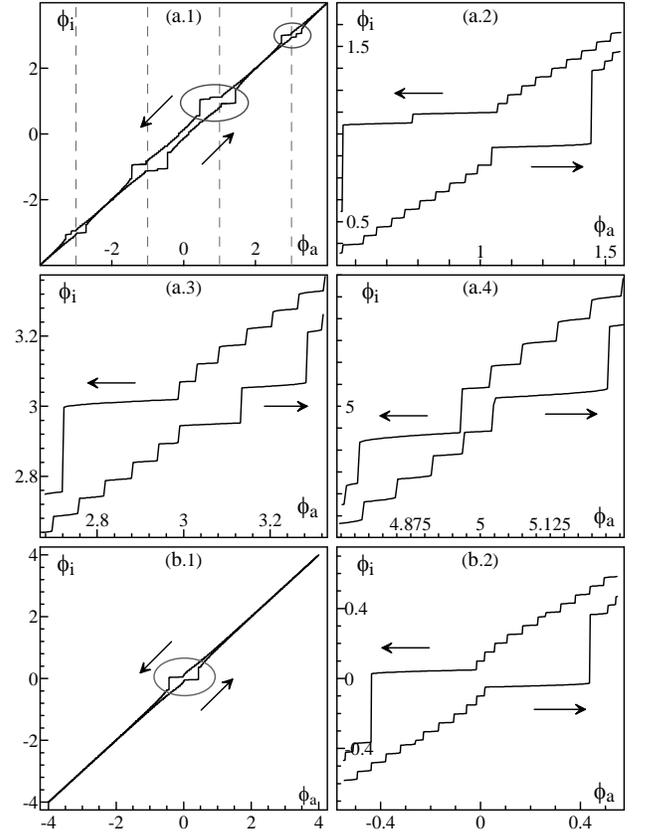}
\caption{(a.1)$\,$-$\,$(a.4) Magnetization, $\phi_i(\phi_a)$, of a
junction with stepwise critical current density and $N=40$; (a.1) the
plateaus with $\phi_i ={\rm const}$ reveal the first four FS states.
The curves $\phi_i(\phi_a)$ in the vicinity of three FS states: (a.2)
$H_a=H_1$, (a.3) $H_a=3H_1$; (a.4) $H_a=5H_1$. (b.1)$\,$-$\,$(b.2)
Magnetization of a junction of conventional superconductors. (b.2)
$\phi_i(\phi_a)$ of a conventional junction in the Meissner
state.
\label{fig_3}}
\end{figure}
\par
Next, we calculate the width of the intervals $\Delta H_a$ of existence
of the FS states by minimizing free energy, ${\cal F}$, of the
junction. This can be done explicitly if we specify the spatial
distribution of the critical current density, $j_c(x)$. Here, for
brevity, we use the one harmonic model, {\it i.e.}, we assume that
$j_c(x)= j_1\sin(2\pi x/l)$:
% 17
\begin{eqnarray}\label{eq_017}
&&{\cal F}=\left({H_a\over H_{\rm sm}}\right)^2-{\Lambda_m\over l}
{H_a\over H_{\rm sm}}{4\pi\phi_i\over\phi_0}+ \nonumber\\
&&\int_0^L\left[\Lambda_m^2\varphi'^2 -{1\over 2}\sin\left({2\pi m\over
l}x +\theta_m\right)\cos\varphi\right]{dx\over L}\,,\ \
\end{eqnarray}
where ${\cal F}$ is normalized by $H_{\rm sm}^2L/8\pi$, with $H_{\rm
sm}=\phi_0/4\pi\lambda\Lambda_m\ll H_m$. If there is no vortices in the
junction, then we have $\psi =0$ and $\varphi(x)=2\pi\phi_i
x/\phi_0l+\varphi_0$, where $\varphi_0$ is a constant. In this case the
minimum value of the free energy, ${\cal F}_m$, is achieved for
$\varphi_0=\theta_m$,
% 18
\begin{eqnarray}\label{eq_018}
{\cal F}_{\rm m}=\!\left({2\pi\Lambda_m\over l\phi_0}\right)^2\!\!\!
(\phi_a-\phi_i)^2 -{\sin\left[\pi N(m-\phi_i/\phi_0)\right]\over\pi
N(m-\phi_i/\phi_0)}\,,\ \
\end{eqnarray}
where $\phi_a=2\lambda lH_a$. Finally, we fix the applied field $H_a$
and minimize ${\cal F}_m$ with respect to the internal flux $\phi_i$.
It follows from this calculation that if $H_a$ is from the narrow
interval $mH_1-H_{\rm sm}<H_a<mH_1+H_{\rm sm}$, then the flux $\phi_i$
is constant and equal to $m\phi_0$, which means that $H_i\approx mH_1$.
\par
% fig. 4
\begin{figure}
\includegraphics[width=0.95\columnwidth]{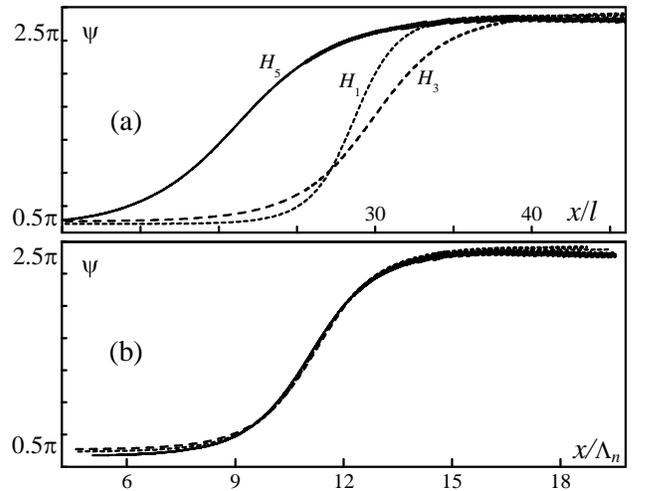}
\caption{(a) The phase $\psi(x)$ distribution for vortices
carrying flux $\phi_0$ at three ``odd'' resonant fields $H_1$, $3H_1$,
and $5H_1$. (b) The same three curves are shown to collapse into one
curve if the coordinate is normalized by $\Lambda_1$, $\Lambda_3$, and
$\Lambda_5$.}
\label{fig_4}
\end{figure}
The above theoretical analysis can be supported by numerical
simulations. To treat both statics and dynamics of the phase difference
$\varphi(x,t)$ we introduce time dependence into Eq. (\ref{eq_008}) and
arrive to
% 19
\begin{equation}\label{eq_019}
\ddot{\varphi} +\alpha\dot{\varphi} -\lambda_J^2\varphi''+
i_c(x)\sin\varphi =0\,,
\end{equation}
where $\alpha\sim 1$ is a decay constant. The second term in the LHS of
Eq. (\ref{eq_019}) describes dissipation. As a result of this
dissipation the system ends up in one of the stable stationary states
which is a solution of Eq. (\ref{eq_008}).
\par
We used the finite-difference explicit method (see for details
\cite{Mints_4}) to solve Eq. (\ref{eq_019}) assuming that the
alternating critical current density $j_c(x)$ is stepwise. The
magnetization curves (dependencies of $\phi_i$ on $\phi_a$) obtained by
numerical simulations are shown in Fig. \ref{fig_3}. The plateaus with
$\phi_i ={\rm const}$ are clearly seen for the series of the four FS
states. The field inside the junction is constant for each of the FS
states in contrast to junctions of conventional superconductors for
which the internal field is constant only in the Meissner state. It is
seen in Fig. \ref{fig_3} that the magnetization curve of a junction in
the FS state is the same as the magnetization curve of a junction of
conventional superconductors but with the field $H_a$ biased by $mH_1$.
Our numerical simulations confirm that the width of the plateau in
which the internal field is constant is proportional to
$\sqrt{|i_m|}=\lambda_J/\Lambda_m$ as it follows from the above
theoretical analysis.
\par
\begin{figure}
\includegraphics[width=0.95\columnwidth]{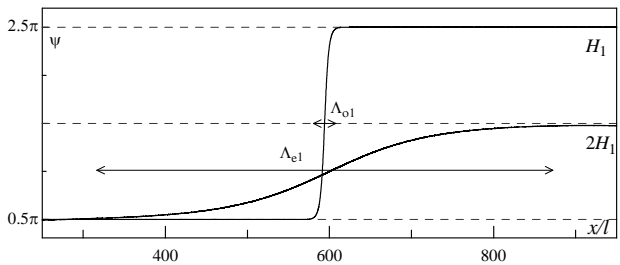}
\caption{The phase $\psi(x)$ distribution for the first ``odd'', $H_a=H_1$,
and ``even'' $H_a=2H_1$ resonant fields describing vortices with fluxes
$\phi_0$ ($H_a=H_1$) and $\phi_0/2$ ($H_a=2H_1$).}
\label{fig_5}
\end{figure}

Next, we simulated numerically the dynamics and statics of single
Josephson-type vortices in the FS states (see Figs. \ref{fig_4} and
\ref{fig_5}). We observed stable Josephson-type vortices carrying
fluxes $\phi_0$ (for ``odd'' resonant fields) and $\phi_0/2$ (for
``even'' resonant fields). The width of vortices carrying flux $\phi_0$
scales proportionally to $\sqrt{|i_m|}$ as follows from Eq.
(\ref{eq_015}). Eq. (\ref{eq_016}) shows that the width of vortices
carrying flux $\phi_0/2$ is by factor $\lambda_J/l\gg 1$ larger than
that of $\phi_0 $.
\par
We found also that periodical motion of single vortices between the
edges of the junction ($x=0$ and $x=L$) produce steps in the IV
characteristics at a series of voltages $V_m=m\phi_0c_s/L$, where $m$
is the number of fluxons inside the junction and $c_s$ is the Swihart
velocity \cite{Tinkham_1}. These high field steps are similar to the
zero field steps in junctions of conventional superconductors
\cite{Fulton_1,Chen_1,Pedersen_1}.
\par
% fig. 5
\par
To summarize, we find new equilibrium FS states in Josephson junctions
with periodically alternating critical current density. The FS states
exist if the applied field is from narrow intervals centered at
equidistant series of resonant fields. If a junction is in FS state,
the flux in the junctions is fixed and the maximum supercurrent across
the junction is significantly enhanced. Two types of single high-field
vortices with flux $\phi_0$ or/and $\phi_0/2$ exist in FS states.
Periodic motion of these vortices between the edges of the junction
results in high field steps in the IV characteristics.
\par
One of the authors (RGM) is grateful to J. R. Clem, V. G. Kogan, J.
Mannhart, and C. W. Schneider for support and stimulating discussions.
\par
\bibliography{FSS}
\end{document}